# ORIGIN: Metal Creation and Evolution from the Cosmic Dawn

J.W. den Herder, L. Piro, T. Ohashi, C. Kouveliotou, D.H. Hartmann, J. Kaastra, L. Amati, M.I. Andersen, M. Arnaud, J-L. Attéia**,** S. Bandler, M. Barbera, X. Barcons, S. Barthelmy, S. Basa, S. Basso, M. Boer, E. Branchini, G. Branduardi-Raymont, S. Borgani, A. Boyarsky, G. Brunetti, C. Budtz-Jorgensen, D. Burrows, N. Butler, S. Campana, E.Caroli, M. Ceballos, F. Christensen, E. Churazov, A. Comastri,  L. Colasanti, R. Cole, R. Content, A. Corsi, E. Costantini, P. Conconi, G. Cusumano, J. de Plaa, A. De Rosa, M. Del Santo, S. Di Cosimo, M. De Pasquale, R. Doriese, S. Ettori, P. Evans, Y. Ezoe, L. Ferrari, H. Finger, T. Figueroa-Feliciano, P. Friedrich, R. Fujimoto, A. Furuzawa, J. Fynbo, F. Gatti, M. Galeazzi, N. Gehrels, B. Gendre, G. Ghirlanda, G. Ghisellini, M. Gilfanov, P. Giommi, M. Girardi, J. Grindlay, M. Cocchi, O. Godet, M. Guedel, F. Haardt, R. den Hartog, I. Hepburn, W. Hermsen, J. Hjorth, H. Hoekstra, A. Holland, A. Hornstrup, A. van der Horst, A. Hoshino, J. in 't Zand, K. Irwin, Y. Ishisaki, P. Jonker, T. Kitayama, H. Kawahara, N. Kawai, R. Kelley, C. Kilbourne, P. de Korte, A. Kusenko, I. Kuvvetli, M. Labanti, C. Macculi, R. Maiolino, M. Mas Hesse, K. Matsushita, P. Mazzotta, D. McCammon, M. Méndez, R. Mignani, T. Mineo, K. Mitsuda, R. Mushotzky, S. Molendi, L. Moscardini, L. Natalucci, F. Nicastro, P. O'Brien, J. Osborne, F. Paerels, M. Page, S. Paltani, K. Pedersen, E. Perinati, T. Ponman, E. Pointecouteau, P. Predehl, S. Porter, A. Rasmussen, G. Rauw, H. Röttgering, M. Roncarelli, P. Rosati, E. Quadrini, O. Ruchayskiy, R. Salvaterra, S. Sasaki, K. Sato, S. Savaglio, J. Schaye, S. Sciortino, M. Shaposhnikov, R. Sharples, K. Shinozaki, D. Spiga, R. Sunyaev, Y. Suto, Y. Takei, N. Tanvir, M. Tashiro, T. Tamura, Y. Tawara, E. Troja, M. Tsujimoto, T. Tsuru, P. Ubertini, J. Ullom, E. Ursino, F. Verbunt, F. van de Voort, M. Viel, S. Wachter, D. Watson, M. Weisskopf, N. Werner, N. White, R. Willingale, R. Wijers, N. Yamasaki, K. Yoshikawa, S. Zane.

J.W. den Herder, E. Costantini, R. den Hartog, W. Hermsen, J. in 't Zand, J. Kaastra, J. de Plaa, P. Jonker, P. de Korte

*SRON Netherlands Institute for Space Research, Sorbonnelaan 2, 3584 CA Utrecht, the Netherlands*

L. Piro, M. Cocchi, L. Colasanti, A. Corsi, A. De Rosa, M. Del Santo, S. Di Cosimo, B. Gendre, C. Macculi, L. Natalucci, P. Ubertini

*INAF - Istituto di Astrofisica Spaziale e Fisica Cosmica, Roma Italy*

T. Ohashi, Y. Ezoe, Y. Ishisaki, h. Kawahara, S. Sasaki

*Tokyo Metropolitan University, Tokyo, Japan*

C. Kouveliotou, A. van der Horst, M. Weisskopf

*Marshall Space Flight Center, Huntsville, USA*

D.H. Hartmann

*Department of Physics and Astronomy, Clemson University, Clemson, USA*

L. Amati, E.Caroli, M. Labanti

*INAF - Istituto di Astrofisica Spaziale e Fisica Cosmica, Bologna Italy*

M.I. Andersen, J. Fynbo, J. Hjorth, K. Pedersen, D. Watson

*Dark Cosmology Centre, Niels Bohr Institute, University of Copenhagen, Denmark*

M. Arnaud

*CEA Saclay, Service d'Astrophysique, Gif-sur-Yvette, France*

J-L. Attéia

*LAT, Observatoire Midi-Pyrénées, Toulouse, France*

S. Bandler, S. Barthelmy, N. Gehrels, R. Kelley, C. Kilbourne, S. Porter, E. Troja,

N. White

*NASA Goddard Space Flight Center, USA*

M. Barbera, G. Cusumano, T. Mineo, E. Perinati, S. Sciortino

*INAF- Istituto di Astrofisica Spaziale, Palermo, Italy*

X. Barcons, M. Ceballos

*IFCA, Santander, Spain*

S. Basa

*Observatoire de Marseille, Marseille, France*

S. Basso, S. Campana, P. Conconi, G. Ghirlanda, G. Ghisellini, D. Spiga

*INAF, Osservatorio Astronomico Brera, Milano, Italy*

M. Boer

*Observatoire de Haute Provence, Haute Provence, France*

E. Branchini, E. Ursino

*Università Roma III, Italy*

G. Branduardi-Raymont, R. Cole, M. De Pasquale, I. Hepburn, R. Mignani, M. Page, S. Zane

*Mullard Space Science Laboratory/UCL, UK*

S. Borgani, M. Girardi, M. Viel

*INAF - Osservatorio Astronomico, Trieste, Italy*

A. Boyarsky

*CERN, Switzerland*

G. Brunetti

*INAF-IRA Bologna, Italy*

C. Budtz-Jorgensen, F. Christensen, A. Hornstrup, I. Kuvvetli

*DNSC/Technical University of Denmark, Copenhagen, Denmark*

D. Burrows

*Penn State University, University Park, Philadelphia, PA, USA*

N. Butler

*University of California, Berkeley, USA*

E. Churazov, M. Gilfanov, R. Sunyaev

*Max-Planck-Insitut für Astrophysik, München, FRG*

R. Content, R Sharples

*Durham University, Durham, UK*

A. Comastri, S. Ettori, L. Moscardini

*INAF – Osservatorio Astronomico Bologna, Italy*

R. Doriese, K. Irwin, J. Ullom

*NIST, Boulder, USA*

P. Evans, P. O'Brien, J. Osborne, N. Tanvir, R. Willingale

*Leicester University, Leicester, UK*

L. Ferrari, F. Gatti

*Istituto Nazionale di Fisica Nucleare, Genova, Italy*

T. Figueroa-Feliciano

*MIT, Cambridge, USA*

H. Finger

*University Space Research Association, Huntsvile, AL, USA*

P. Friedrich, P. Predehl, S. Savaglio

*Max-Planck-Institut für Extraterrestrische Physik, Garching, FRG*

R. Fujimoto, A. Hoshina

*Kanazawa University, Kanazawa, Japan*

M. Galeazzi

*University of Miami, USA*

P. Giommi

*ASI Data Center, Italy*

J. Grindlay

*CfA, Harvard, Cambridge, USA*

O. Godet, E. Pointecouteau

*CESR Centre d'Etude Spatiale des Rayonnements, Toulouse France*

M. Guedel

*University of Vienna, Vienna, Austria*

F. Haardt, R. Salvaterra

*University of Insubria, Como, Italy*

H. Hoekstra, H. Röttgering, J. Schaye, F. van de Voort

*Leiden University, Leiden, the Netherlands*

A. Holland

*Open University, Milton Keynes, UK*

T. Kitayama

*Toho University, Chiba, Japan*

Y. Suto

*University of Tokyo, Tokyo, Japan*

N. Kawai

*Tokyo Institute of Technology, Japan*

A. Kusenko

*University of California at Los Angeles, CA, USA*

R. Maiolino, F. Nicastro

*INAF Osservatorio Astronomico di Roma, Rome, Italy*

M. Mas Hesse

*Centro de Astrobiología (CSIC-INTA), Madrid, Spain*

K. Matsushita, K. Sato

*Tokyo University of Science, Tokyo, Japan*

P. Mazzotta

*Universitá de Roma Tor Vergata, Rome, Italy*

D. McCammon

*University of Wisconsin, Madison, WI, USA*

M. Méndez

*Groningen University, Groningen, the Netherlands*

K. Mitsuda, Y. Takei, T. Tamura, M. Tsujimoto, N. Yamasaki

*Institute of Space and Astronautical Science, JAXA, Kanagawa, Japan*

R. Mushotzky

*University of Maryland, College Park, MD, USA*

S. Molendi, E. Quadrini

*INAF – Istituto di Astrofisica Spaziale e Fisica Cosmica, Milano, Italy*

F. Paerels

*Columbia University, NY, USA*

S. Paltani

*ISDC, University of Geneva, Versoix, Switzerland*

T. Ponman

*University of Birmingham, Birmingham, UK*

A. Rasmussen, N. Werner

*KIPAC/Stanford, Palo Alto, CA, USA*

G. Rauw

*Liege University, Liege, Belgium*

O. Ruchayskiy, M. Shaposhnikov

*Ecole Polytechnique Fédérale de Lausanne, Switzerland*

M. Roncarelli

*University of Bologna, Bologna, Italy*

P. Rosati

*ESO, Garching, FRG*

K. Shinozaki

*Aerospace, Research and Development Directorate, JAXA, Ibaraki, Japan*

M. Tashiro

*Saitama University, Saitama, Japan*

Y. Tawara, Y. Furuzawa

*Nagoya University, Aichi, Japan*

T. Tsuru

*Kyoto University, Kyoto, Japan*

F. Verbunt

*Utrecht University, Utrecht, the Netherlands*

S. Wachter

*Caltech, California, USA*

R. Wijers

*University of Amsterdam, the Netherlands*

K. Yoshikawa,
*Tsukuba University, Ibaraki, Japan*

# Abstract

ORIGIN is a proposal for the M3 mission call of ESA aimed at the study of metal creation from the epoch of cosmic dawn. Using high-spectral resolution in the soft X-ray band, ORIGIN will be able to identify the physical conditions of all abundant elements between C and Ni to red-shifts of $z$=10, and beyond. The mission will answer questions such as: *When were the first metals created? How does the cosmic metal content evolve? Where do most of the metals reside in the Universe? What is the role of metals in structure formation and evolution?* To reach out to the early Universe ORIGIN will use Gamma-Ray Bursts (GRBs) to study their local environments in their host galaxies. This requires the capability to slew the satellite in less than a minute to the GRB location. By studying the chemical composition and properties of clusters of galaxies we can extend the range of exploration to lower redshifts ($z\sim 0.2$). For this task we need a high-resolution spectral imaging instrument with a large field of view. Using the same instrument, we can also study the so far only partially detected baryons in the Warm-Hot Intergalactic Medium (WHIM). The less dense part of the WHIM will be studied using absorption lines at low redshift in the spectra for GRBs.

The ORIGIN mission includes a Transient Event Detector (coded mask with a sensitivity of 0.4 photon/cm$^2$/s in 10 s in the 5-150 keV band) to identify and localize 2000 GRBs over a five year mission, of which ~65 GRBs have a redshift >7. The Cryogenic Imaging Spectrometer, with a spectral resolution of 2.5 eV, a field of view of 30 arcmin and large effective area below 1 keV has the sensitivity to study clusters up to a significant fraction of the virial radius and to map the denser parts of the WHIM (factor 30 higher than achievable with current instruments). The payload is complemented by a Burst InfraRed Telescope to enable onboard red-shift determination of GRBs (hence securing proper follow up of high-z bursts) and also probes the mildly ionized state of the gas. Fast repointing is achieved by a dedicated Controlled Momentum Gyro and a low background is achieved by the selected low Earth orbit.

# Introduction

Metals play a very important role in star formation and stellar evolution, and ultimately lay the foundation of planet formation and the development of life. Beginning with metal free (Population III) stars, the cycle of metal enrichment started when their final explosive stages injected the first elements beyond Hydrogen and Helium into their pristine surroundings. These ejecta created the seeds for the next generation of stars (Population II). So the cycle of cosmic chemical evolution began. Baryons trapped in dark matter potential wells started to form stars, which then became the building blocks of proto-galaxies. Galaxies merged, building up massive structures, and provided the energetic radiation that re-ionized and lit up the dark Universe, starting the cosmic dawn. Finding out when these first stars were created and how the metal abundances evolved is the core quest of ORIGIN. Models of hierarchical structure formation suggest that these early proto-galaxies had low masses, small luminosities, and were metal poor. Simulations of star formation in these environments indicate that the typical mass of the earliest stars exceeded a hundred solar masses. These processes started a few hundred million years after the Big Bang. Identifying objects at these look back times is a frontier of observational cosmology. Ultra-deep surveys in the optical and the infrared have resulted in a few detections already out to redshifts of ~10 (about 500 million years after the Big Bang), but it is clear that star formation started even earlier. The only natural phenomena that can *directly* probe the baryonic environments of these first stars are Gamma Ray Bursts (GRBs). These ultra-luminous explosions are believed to be embedded in star-forming regions, effectively acting as beacons that illuminate and pinpoint the unique cradles of early nucleosynthesis.

Following these early phases, gravity leads from the first population of stars in small galaxies to large-scale clusters of galaxies. The production and distribution of elements within these evolving dynamic structures needs to be mapped in greater detail than has been achieved to date. Simulations of large-scale structure by Springel et al. (2006), and others, beautifully demonstrate how the power spectrum of matter evolves, and how voids and filaments emerge as natural features of the dynamic Universe. Simulations of Dark Matter (DM) on the scale of the Milky Way's halo (Diemand et al. 2008) show that the DM dominated local environment is highly structured as well. Baryons, making up only about 4% of the cosmic budget, trace some of these structures, and are essential to our ability to gather knowledge of how and when the Universe produces stars and gaseous flows. Only a fraction of the baryons end up in stars, most remain in diffuse structures; and some baryons that did end up in stars, re-emerge in the diffuse component after nuclear processing in the explosive final stages of massive stars. Transport of gas within galaxies, i.e., infall into and outflow from galaxies, and similar processes on the scale of clusters, created a rich distribution of metal-enriched gas on small and large scales. We need an accounting of the whereabouts and the conditions of the cosmic baryons to understand the feedback processes responsible for the density-temperature-abundance patterns that are observed in galaxies, clusters of galaxies, and the filamentary bridges between clusters predicted in simulations. Extending our knowledge about the metal enrichment processes on large scales is the second quest of ORIGIN. The study of chemical composition in clusters as a function of redshift will help us understand the conditions under which these structures were formed. In the local Universe the metal content of the WHIM can be used to distinguish between different metal diffusion models.

The study of how metal enrichment proceeded from the initial (primordial) conditions emerging from Big Bang Nucleosynthesis to those of present stars and galaxies is a formidable challenge, but is now possible for the first time due to advances in cryogenic X-ray detectors. *X-ray spectroscopy has the unique capability of simultaneously probing all the elements (C through Ni), in all their ionization stages and all binding states (atomic, molecular, and solid), and thus provides a model-independent survey of the metals*. ORIGIN employs high-resolution X-ray spectroscopy and imaging to detect most of these elements to very high redshifts. The most distant star forming regions can be probed with rapid response spectroscopy of bright GRBs and local cluster structures can be studied with a wide field of view imaging survey. The filamentary WHIM is probed along the sight lines of the GRBs in absorption and can also be mapped using the wide field of view imaging capability of ORIGIN. Infrared spectroscopy will provide independent and complementary information on chemical abundances by detecting low ionization absorption lines. Thus a combination of near-field cluster surveys with far-field GRB response-spectroscopy provides an optimal strategy to map cosmic chemical evolution from re-ionization to the present.

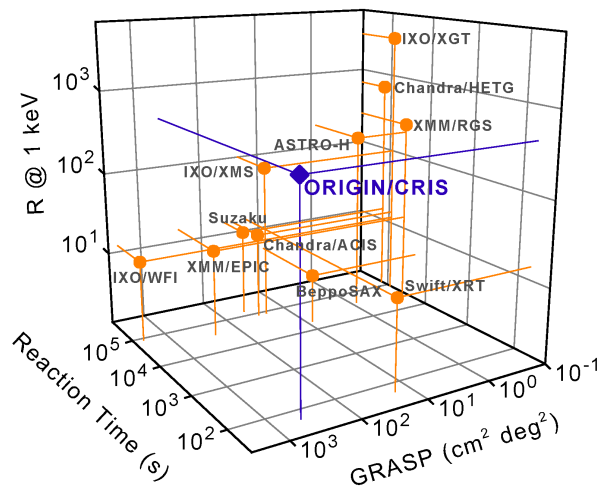

**Fig 1** The key characteristics of ORIGIN (grasp, reaction time, resolution) compared to existing and proposed missions.

Figure 1 demonstrates the observational space uniquely address by ORIGIN. The combination of fast re-pointing response with the high spectral resolution (R) and the large grasp (Area x Field of View), make ORIGIN a powerful tool for *transient and wide field high-resolution spectroscopy*. In contrast to previous GRB-dedicated satellites (e.g. *Swift*), ORIGIN will be totally autonomous in determining not only the location, but also the redshift, physics and chemistry of the ISM surrounding the GRB. Compared to the ASTRO-H mission, a Japanese mission to be launched in 2014, ORIGIN has, in addition to the fast response, a larger field of view (factor of 100) and a larger effective area below 2 keV (factor of 7), combined with a factor of 2 better spectral resolution. ORIGIN is highly complementary to the capabilities of the International X-ray Observatory (IXO), which has higher angular resolution and effective area.

# Science

With its unique capability of high spectral resolution and imaging, ORIGIN will advance many fields of astrophysics. Its continuous monitoring of a large part of the sky will further increase the scientific return, as ORIGIN is sensitive to all types of transient phenomena in the hard X-ray band. In this section, we limit ourselves, however, to the main quests of ORIGIN: the study of the cosmic metal enrichment history.

## The first stars

ORIGIN is uniquely designed to answer several key questions about star forming processes in the early Universe. By measuring GRB redshifts and abundances in the circumburst medium deep into the era of re-ionization ($z>6$), we will discover when star formation started and how it evolved into the present day structures. Most importantly, ORIGIN is the only Cosmology mission that can chart the abundance patterns that prevailed in these early dark matter – baryon systems.

GRB explosions taking place in these proto-galaxies would be easily detected with ORIGIN. Figure 2 shows a simulated X-ray afterglow of an explosion at $z=7$, as measured with the ORIGIN/CRIS. Multiple narrow spectral lines are identified with ionized metals in the burst environment, allowing measurement of relative abundances and the determination of the redshift. To secure redshift measurements even in the case of exceedingly low metallicities, the mission also carries an IR telescope to measure the Lyman break in GRB spectra. This IR telescope provides also column densities for low ionization states of elements.

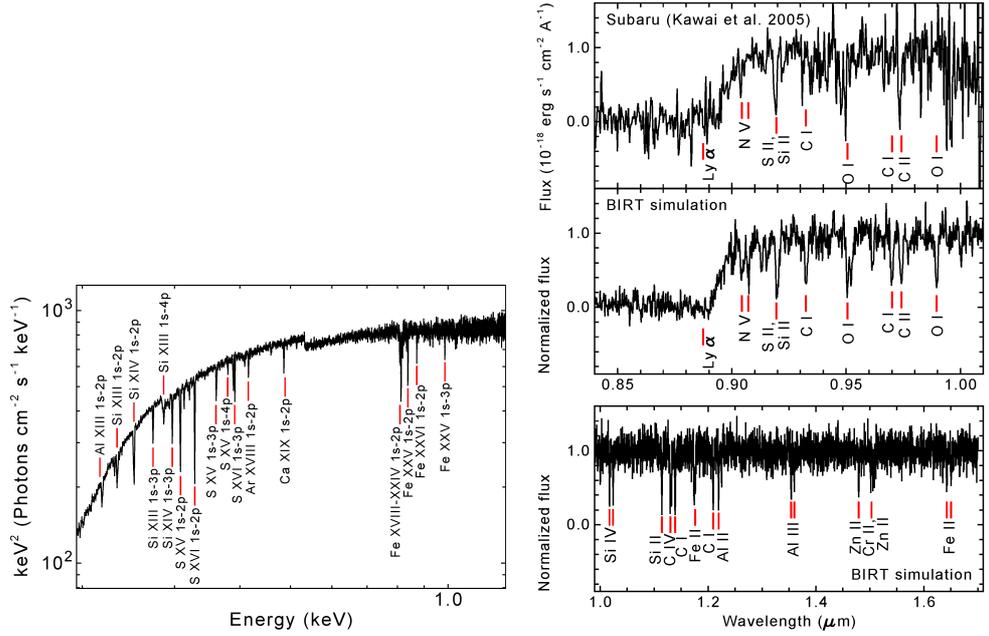

**Fig 2.** Left: simulated CRIS X-ray spectrum of a medium bright afterglow ($S_X=4\times10^{-6}$ erg/cm$^2$ integrated between 50 s – 50ks) at $z=7$ characterized by deep narrow resonant lines of Fe, Si, S, Ar, Mg, from the gas in the environment of the GRB. An effective column density of $2\times10^{22}$ cm$^{-2}$ has been adopted, consistent with the values observed in GRB afterglows. Right top panel: IR spectrum of GRB050904 at $z=6.3$ as observed with Subaru 3.4 days after the burst (Kawai et al. 2006). Middle and Bottom panels: simulation of the same spectrum with BIRT, starting 280 s after the GRB trigger and lasting for 1000 s. Only a part of the BIRT wavelength range is shown. The Lyman break is very well identified in the spectrum as well as most of the key absorption lines.

ORIGIN will collect about 400 GRBs per year covering the full redshift distribution. About twice per month a GRB from the re-ionization era will trigger the instruments. The resulting multi-element abundance patterns will map the evolving chemical composition of the early Universe, "fingerprint" the elusive Pop III stars, and constrain the shape of the Initial Mass Function (IMF) of the first stars.

## The history of metal production in clusters of galaxies

The cosmic history of metal production and the circulation of these metals throughout the Universe is a fundamental astrophysical question. Clusters of galaxies are excellent laboratories to study these processes since 85% of the baryons are in the hot X-ray emitting gas and, due to their deep gravitational potential, clusters retain all the metals that were produced inside them. High-resolution X-ray spectroscopy of this gas will unveil the history of nucleosynthesis (figure 3).

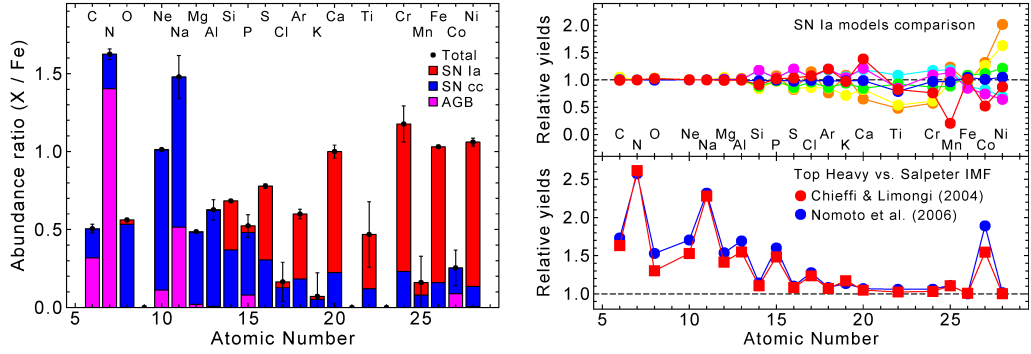

**Fig 3**. Left: Expected abundance ratios relative to iron for a 100 ks exposure with ORIGIN for a typical cluster, showing the contributions from AGB stars, SN Ia an SNcc (cf. Werner et al. 2006; De Plaa et al. 2007). The relative weight of low and high Z elements varies considerably for the different contributions. ORIGIN will measure 20 elements from C to Ni (XMM-Newton can only measure ~7 elements). Right: total metal yields compared to a standard model. Top panel: Variations .for different SN Ia models. Bottom panel: Top Heavy IMF compared to the Salpeter IMF. ORIGIN will discriminate these SN Ia models and the IMF by accurately measuring the relative metal content.

Almost all metals heavier than oxygen are produced by supernova (SN) progenitors and most of the atoms heavier than silicon originate from the SNIa sub-class. It is still unknown which progenitor systems produce all, or the bulk of Type Ia explosions. Analysis of the chemical abundance of clusters with CCD detectors on XMM-Newton and Suzaku have shown an abundance pattern of Si, S, Ca, Ar, Fe and Ni which is not consistent with the theoretical predictions of classical SNIa. This indicates either that the theory needs modification, as detailed analysis of the brightest SNIa remnant (Tycho) indicates, or that clusters are predominantly enriched by different types of SNIa. However, these data have only been obtained for the very brightest, local clusters and even then only in their central regions. It is thus not clear how general these results are. ORIGIN measures a much wider range of clusters over a large redshift range, and determines the precise ratios of elemental abundances such as Ca to Ar and Ni to Fe which are sensitive to the details of the explosion mechanism. Thus the ORIGIN X-ray spectra will constrain the nucleosynthesis models for SNIa (see figure 3 left panel for typical abundance ratio's). In addition to the abundant elements ORIGIN will measure trace elements like Na, Al, Ti, Cr, Mn and Co, integrated over the cluster core. The Mn/Cr ratio is a sensitive tracer of the metallicity of the progenitor, while the Na abundance is a sensitive measure of the slope of the IMF. Thus, measurements of the abundances of these elements reveal the epoch when these systems were formed and their IMFs. The bulk of the lighter metals (O, Ne, Mg) are formed by core-collapse supernovae (ccSNe or SNII), although these systems also produce heavier elements. To disentangle both contributions, the full range of elements from O to Ni needs to be measured. SNII have massive, short-lived progenitors, so the bulk of the metals created by them is produced and redistributed rapidly after star formation, starting at the epoch of re-ionization and peaking around $z$=2. However, details about the IMF and what these stars exactly produced are not well known (Bastian et al. 2010). Measuring the abundances of N-Mg will constrain these parameters (figure 3). Finally, C and N have a different origin in intermediate-mass AGB stars, and are returned to the ISM by stellar mass loss. When and where these metals were produced is uncertain, and ORIGIN will be able for the first time to map them both in space and time.

## Evolution of clusters of Galaxies

While the *integrated* time history of metal enrichment can be studied in detail in a sample of relatively nearby bright clusters and groups of galaxies, their evolution can also be studied directly by observing clusters as a function of redshift. Until now, it has only been possible to measure evolution of the Fe content (see figure 4). With ORIGIN, we will obtain accurate abundances of many key elements, including iron (mostly from SNIa), oxygen (mostly from ccSNe) and nitrogen (predominantly AGB stars) out to cluster redshifts of 1.3, 1.0, and 0.8, respectively. We will also obtain abundances of several other elements and will address the following questions: How do the abundances in clusters evolve over time? Did ccSNe dominate at early stages, or was there a more complex evolving population? Is the AGB star population co-evolving with the SNIa population?

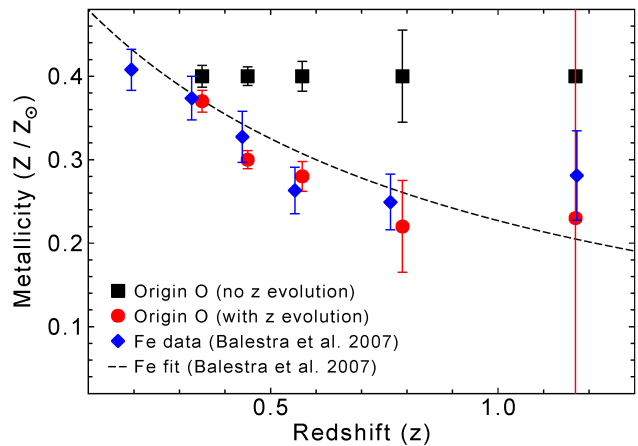

**Fig. 4** Time evolution of the iron abundance in a sample of high-z clusters as measured with Chandra. Each point is the average for ~10 clusters (Balestra et al. 2007). ORIGIN will do similar measurements out to $z>1.2$ for O (as shown), Ne, Mg, Si, S and Fe and for 5 other elements up to $z$~0.5.

## Cosmic filaments

Cosmological hydrodynamic simulations suggest that the missing baryons at $z<2$, contributing ~40% of the cosmic baryon budget, can be accounted for by a diffuse, highly ionized WHIM, preferentially distributed in large-scale filaments connecting the nodes of the Cosmic Web. This gas is extremely hard to detect: its bulk resides in structures with $T>10^6$ K but the thermal continuum emission is much too faint to be detectable against the overwhelming fore - and background emission. The only characteristic radiation from this medium will be from discrete transitions of highly ionized C, N, O, Ne, and possibly Fe. At $T≥10^6$ K, *the primary tracers (O VII, O VIII) cannot be probed with UV observations and are only detectable by soft X-ray spectroscopy*. Indications of the WHIM in this temperature range were obtained with Chandra and XMM-Newton observations of the 21.6 Å resonance absorption line of OVII in the sight line of the Sculptor Wall. Evidence for the warm tail of the WHIM, where 10-15% of the missing baryons reside, has been obtained via UV-absorption line studies with FUSE and HST-COS. ORIGIN will have sufficient line sensitivity and energy resolution to measure gas densities down to $10^{-5}$ cm$^{-3}$, ~30 times smaller than currently probed in clusters. ORIGIN can detect these WHIM lines both in emission, and in absorption against early-stage GRB afterglows, which are the only sufficiently bright, numerous, and distant ($z>1$) sources to guarantee a statistically significant sample size of WHIM detections. An example of what GRB X- ray spectroscopy would yield is shown in Figure 5. We modeled the properties of the WHIM with large scale DM + hydrodynamic simulations, with a parameterized treatment of stellar feedback, using the simulations of Borgani et al. (2004) and Viel et al. (2004) to investigate various WHIM models.

While mere detections of X-ray absorption and emission from ionized metals will reveal the presence of the 'missing baryons', our more ambitious goal is to detect and characterize the physical state of the WHIM: its temperature, density, spatial distribution, and trace the metal enrichment of the IGM and its interplay with the history of star formation and feedback processes. Figure 6 illustrates how well ORIGIN will discriminate among different IGM metal enrichment models through absorption spectroscopy of the WHIM. In absorption we expect to measure about 300 filaments in 5 years from a sample of 500 bright afterglows (Branchini et al. 2009). These observations will allow us to estimate the temperature and the density of each absorption system with ~15% accuracy. In addition, for about 30% of these systems we will also detect the associated X-ray line emission, once the afterglow has faded. A deep field of several square degrees will secure faint emission structures over 16 Mpc at $z=0.3$. Intensities of emission line (scales as $n^2L$) and absorption lines ($nL$) from the same WHIM cloud will yield the density $n$ and the line of sight depth $L$ in a very effective way.

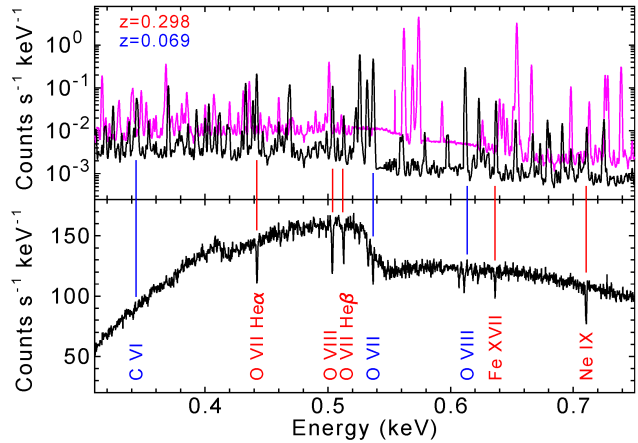

**Fig. 5** Emission spectrum of a 4 arcmin$^2$ area (top), and absorption spectrum (bottom) of the same region of the sky, as measured by ORIGIN. The top panel shows the emission of two red-shifted components in black, while the emission of the Galactic foreground is displayed in purple. The bottom panel shows the spectrum of the same systems, but now in absorption using a bright GRB afterglow as a beacon.

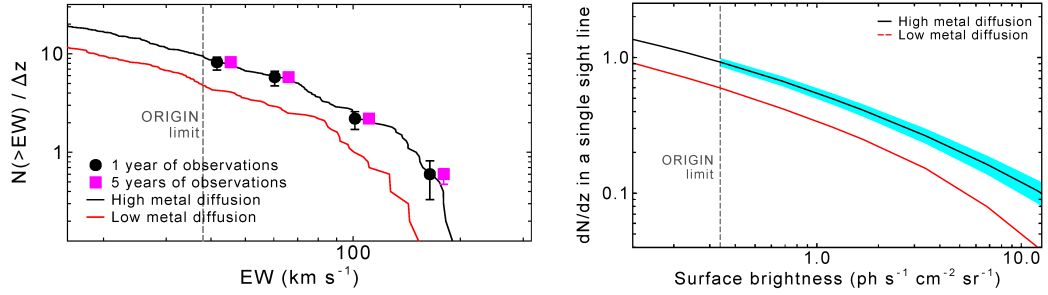

**Fig. 6** left: filaments in absorption. Solid lines show predicted number of OVII absorption lines per unit redshift as a function of line EW. Black bars show the OVII line statistics built up in 1 yr and magenta in 5 years with ORIGIN. The curves are for different metal diffusion models: highly localized to the region of metal synthesis (red), and diffusion into the IGM via SN feedback-related processes (black). These two models represent the spread in current theoretical predictions and can, already after a year, be clearly distinguished. Right: filaments in emission, showing the expected number of O VII+O VIII lines per unit redshift above a given O VII surface brightness. The cyan area gives the observational uncertainties assuming the high-metal diffusion model. It illustrates that the measurements are very distinctive. The case refers to a 50"x50" field, observed for 1 Ms with ORIGIN.

A 1 Ms exposure with ORIGIN will reveal about 1200 emitting systems per $deg^2$ with joint $5\sigma$ detection of O VII and O VIII line, and will measure the temperature with ~30% error (Takei et al. 2010). Figure 6 shows the expected dN/dz of these emitters. The observation of so many different WHIM properties will allow discrimination among different stellar feedback and metal diffusion mechanisms. Combining constraints from different observations lifts possible model degeneracy.

## Mission profile

To enable the science described above we use a Transient Event Detector (TED) to locate GRBs and fast repointing to observe the GRB afterglow with a wide field X-ray telescope equipped with a Cryogenic Imaging Spectrometer (CRIS) alongside a capable Burst InfraRed Telescope (BIRT). The wide field of the X-ray instrument also provides sensitive complimentary observations of the local Universe. This powerful combination gives us three different tracers of chemical evolution covering three different epochs.

Absorption lines from the environment of GRBs exhibit Equivalent Widths of ~1 eV, while even weaker absorption lines at EW ~0.2 eV are imprinted by the cosmic filaments. Detection of such weak lines requires a fluence S>500 photons $eV^{-1}$, corresponding to $10^{-6}$ erg $cm^{-2}$ in the 0.3–10 keV band. This requirement naturally implies an effective area A>1000 $cm^2$ and a spectral resolution $\Delta E \leq 2.5$ eV, which is provided by the *Cryogenic Imaging Spectrometer* (CRIS). Because GRB afterglows fade quickly, one needs *rapid localization and repointing* capability, with a spectrometer pointing at the source within 60 s after the trigger. A *Transient Event Detector* (TED) with FoV of ~4 sr and a sensitivity of 0.4 photon $cm^{-2}$ $s^{-1}$ between 5–150 keV, integrated over 10 s, is required to detect at $12\sigma$, and thus localize within 3', 2000 GRBs in 5 years. From the prompt and afterglow fluence distributions observed by Swift we expect ~500 afterglows with fluence >$10^{-6}$ erg $cm^{-2}$, sufficient to carry out high-resolution spectroscopy. Out of the 2000 GRB, TED will provide *~125 GRBs at z>6 and 65 at z>7* over the mission life time. This sample allows us to derive quantitative conclusions. For the brightest afterglows (fluence >$10^{-6}$ erg cm$^{-2}$ in the 0.3–10 keV band), we can measure metal column densities as low as H equivalent $10^{21}$ cm$^{-2}$. This will allow us to access gas at metallicities as low as 1% of solar for the denser regions expected in early stars; in even denser regions ($N_H$>$10^{23}$ cm$^{-2}$) the accuracy will be further improved. The redshift of these afterglows will be measured with a precision of ~ 0.1%. For regions of very low metallicity, the redshift will be secured by measuring the Lyman break. Therefore *a Burst Infra-Red Telescope* (BIRT) complements the payload. With a resolution of 20 over the range of 0.5–1.7 μm, it allows the determination of the redshift of all observed bursts between 5.5 and 12 using the Lyman break in the spectra within 1%. With an additional resolution of 1000, this instrument can also measure low ionization lines, complementing the characterization of metallicity.

ORIGIN will complement its study of the high-z Universe with studies of the metal content at lower redshifts (clusters of galaxies, WHIM). This requires a large field of view and equally good energy resolution. Here, spectral resolution is set by the required contrast of the extraordinarily faint emission against the background (instrument background, unresolved extragalactic point sources, galactic foreground emission and, in the case of clusters, thermal continuum emission). The *low background*, crucial for these measurements, is achievable with a Low Earth Orbit (LEO), a small focal ratio of the telescope, and optimized detector shielding. The instrumental background will be as low as 2 $10^{-5}$ counts s$^{-1}$ arcmin$^{-2}$ keV$^{-1}$, which is about an order of magnitude lower than the level of the cosmic X-ray background. The angular resolution of the CRIS is set by the typical size of gas concentrations in WHIM filaments of about 30" (~100 kpc at z=0.2), the field of view and the capability to spatially resolve the emission (e.g. of groups of galaxies, turbulent velocity in clusters, etc.). With a 30" HEW and a 30' field of view we will be able to observe these objects in a single or a few adjacent observations. The expected characteristic emission line intensity of the WHIM is ~0.1 photon cm$^{-2}$ s$^{-1}$sr$^{-1}$ in the strongest O K-shell line (out to redshift 0.3). CRIS will detect such an emission from a typical filament at the 5σ level easily in 1 Ms.

Finally, the fast repointing capability will allow ORIGIN to measure *the WHIM filaments in absorption* using GRBs as backlights (for ~250 line of sights). Combined with measurements of these same filaments in emission, the density of the filament can be uniquely determined.

The prime goal of the IR telescope (BIRT) is to enable the determination of the redshift for GRBs at *z*>5.5 irrespective of the metallicity. This is achieved with low-resolution spectra for GRBs over the range 0.7–1.7 μm corresponding to the redshifted Lyman break. The collecting area should be such that a large fraction of all GRBs will provide a good detection. BIRT, with a limiting magnitude of $H_{AB}$ = 22 in imaging mode, covers the known decay of GRBs observed by Swift and previous missions. Obscured bursts, which will escape the redshift determination by BIRT, will have large column densities and therefore their redshifts can be determined by CRIS.

## Detection of high redshift GRBs

GRBs are amongst the best sources to study the high redshift Universe, due to their existence at high redshift, combined with their exceptional brilliance (cf. galaxies and QSOs) and lack of proximity effects (cf. QSOs). The afterglow flux of high-$z$ events is comparable to those of closer bursts, due to the effect of spectral K-correction and because time dilation leads to sampling of the earlier, brighter part of the afterglow.

The TED sensitivity, low energy threshold and field of view were optimized in order to localize high-$z$ bursts. The low energy threshold, compared to *Swift*, has the double advantage of increasing the sensitivity and bringing into the instrument bandpass more high-z events, whose peak energy is redshifted. It also makes the instrument sensitive for X-ray flashes. A larger solid angle increases the number of events. In the trade-off between the sensitivity and solid angle we have favored the latter. This choice results in a larger number of (high-$z$) events characterized by afterglows bright enough to derive a redshift on the fly. Thus the four modules of the TED are oriented in different directions, providing a total field of view of 4 sr. The expected number of high-$z$ bursts has been calculated based on the independent models described in Salvaterra et al. (2010) and Butler et al. (2010). Both models reproduce the observed log$N$-log$S$ relation. They also reproduce the number of high-z bursts ($z$>5) observed by *Swift*, taking into account that this number is a lower limit, given the observational bias against the identification of high redshift bursts. Most models indicate that the rate of GRBs increases faster than the SFR at high redshift.

By folding into Butler's model the TED performance we expect to detect about 400 bursts per year, 25 at $z$>6 and 13 with $z$>7. From the model of Salvaterra et al. (2010) we derive ranges which are consistent with the model of Butler et al. (2010). If we take the most pessimistic estimate we derive a lower limit of 12 GRB at $z$>7 in 5 years. These numbers are consistent with the fraction of high redshift ($z$>6) bursts estimated from samples of optical-IR follow up of *Swift* bursts (Jakobsson et al. 2006, Greiner et al. 2010). TED will deliver a factor 4 more bursts than Swift due to its decreased low energy threshold and increase in solid angle (see figure 7).

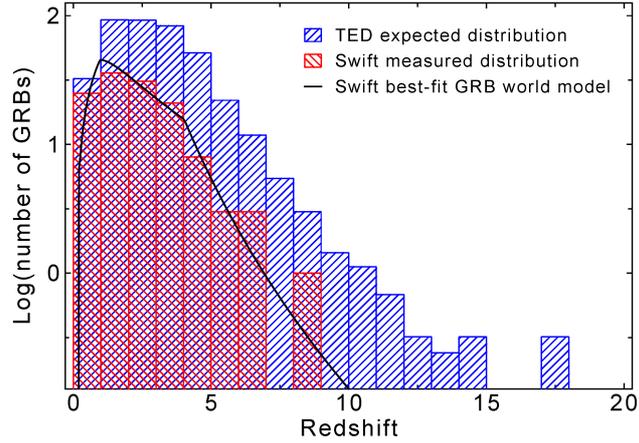

**Fig. 7** The expected distribution in redshift of events localized by TED in one year. The red histogram and corresponding curve are for *Swift* (Butler et al. 2010). Using the same model the ORIGIN expectations are shown in blue.

## Observation program

The observation program covers 4 key topics: (a) when a GRB is detected the satellite will slew to this position and determine the redshift in <2 ks using BIRT. Bright or high redshift GRBs will be observed for a total of 50 ks. For the bright GRBs this is sufficient to collect typically $10^6$ counts in the X-ray spectrum. For these sources BIRT will provide photometry and a high resolution spectrum (R=1000). These events will be used to study GRB host galaxies. At the same time bright GRBs are used as backlight to detect the filaments of the cosmic web in absorption. A total of 2000 GRBs over 5 year are expected, of which we observe 500 bright GRBs for 50 ks; (b) to study metal abundances in the nearby Universe observations of different clusters is planned. These observations cover abundance patterns inside clusters, the metal content up to a significant fraction of their virial radii and the evolution of clusters with redshift. A total of 20 Ms is needed for this; (c) the present Universe will be studied by a deep map of 2.5x2 $deg^2$ and we expect with the given sensitivity to characterize the denser part of filaments ($\rho_{gas}/<\rho_{gas}> >80$) in the O VII and O VIII lines. In addition there will be a guest observer program (~ 30% of the time) which allows for the community to exploit the unique capabilities of ORIGIN.

## Instruments

The mission requirements correspond to a payload with 3 instruments: The *Cryogenic Imaging Spectrometer* (CRIS) is the prime instrument. It allows for wide field imaging of a 30' area on the sky with a spectral resolution <2.5 eV (inner array) and <5 eV (outer array). Its effective area is >1500 $cm^2$ at 1 keV and >150 $cm^2$ at 6 keV. A separate section of the detector has been optimized for the detection of the very high count rates expected from GRB afterglows. Its confusion limit is $10^{-15}$ erg $cm^{-2}s^{-1}$ for 0.5–2 keV and its point source line detection sensitivity at 0.5 keV (5σ) is typical $2\ 10^{-7}$ photons $cm^{-2}\ s^{-1}$ for a 100 ks observation. The *Transient Event Detector* (TED) will detect GRBs in the 5–150 keV energy range, similar to the Swift/BAT. Its solid angle is >4 sr and its sensitivity is >0.4 photons/$cm^2$/s in the 5–150 keV range. The *Burst InfraRed Telescope* (BIRT) has the prime goal to determine the redshift for all bursts beyond $z>5.5$, irrespective of the metallicity of the host galaxy. It has a bandpass of 0.5–1.7 µm, an H-band sensitivity limit $H_{AB}$=20.8, and a field of view of 6'x6' with a low-resolution spectrometer (prism, mode LOW-RES). We will implement two additional BIRT modes: a photometric mode (IMAGE) with a sensitivity limit of $H_{AB}$=22.2, which allows an accurate determination of positions in four different bands; and a high resolution mode (HIGH-RES) with R~1000 to derive column densities. The relevant absorption lines yield redshifts for GRBs at $z$<5.5. The three instruments work together for GRBs as illustrated in Figure 9.

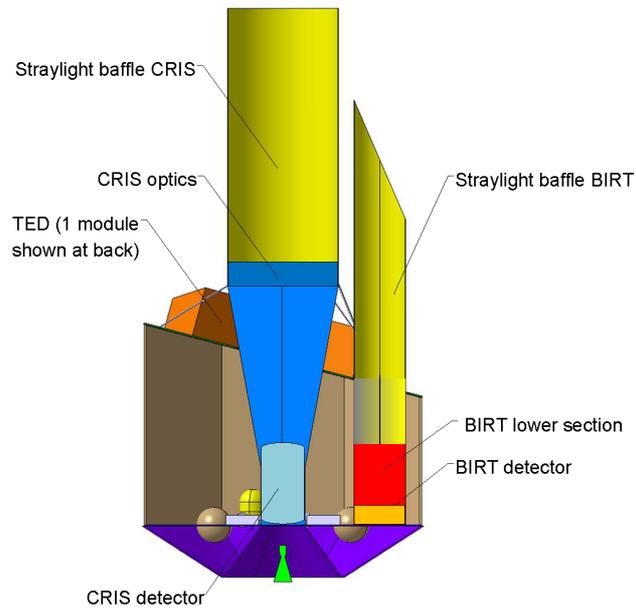

**Fig. 8** The three ORIGIN instruments including their main components. One TED unit at the back is visible (orange). The lower section of BIRT contains the primary and secondary mirror.

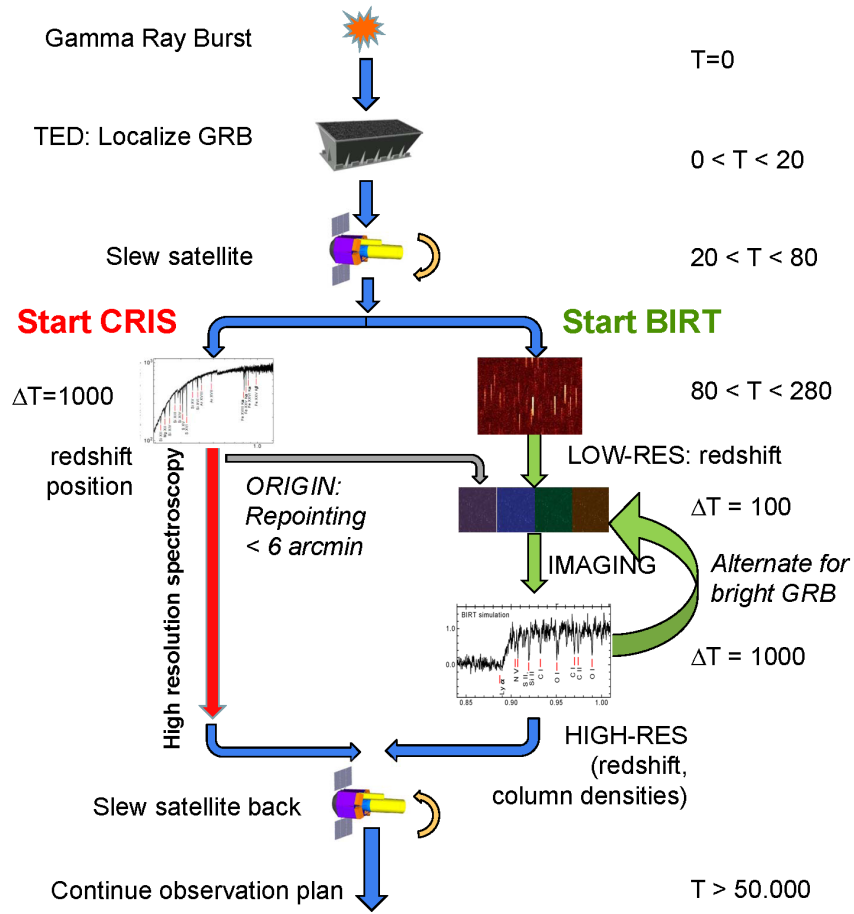

**Fig. 9** Timeline of ORIGIN observations following a GRB detection. Within 200 seconds the redshift of the burst is determined by BIRT.

Bright GRBs ($>10^{-6}$ erg/cm$^2$ in the 0.3–10 keV band) will be localized coarsely (<6') by TED. The spacecraft will slew rapidly and observations by CRIS and BIRT commence. CRIS will use its GRB spectrometer section to obtain a sub-arcmin position, while BIRT will use its LOW-RES spectrometer to obtain a rapid redshift. A shorter slew will then allow use of the BIRT IMAGE mode to identify a sub-arcsec position, while the wide field of the CRIS spectrometer will allow X-ray data collection to continue. BIRT will also obtain spectra for bright GRBs or fainter GRBs at low redshifts: position information from the IMAGE mode of BIRT is fed to a small tip/tilt mirror (<1') which redirects the beam onto the BIRT spectrometer. The same mirror, in combination with a dedicated gyro attached to BIRT, is also used to correct for high frequency jitter in the BIRT pointing. BIRT will alternate between the IMAGE and HIGH-RES mode. If the source fades below a certain brightness (set from the ground), the GRB observation will end and the normal observing program is resumed. During this process TED continues to monitor the sky for transient events, however, depending on user criteria, a GRB observation can be prematurely ended to follow up a more exciting event.

## The Cryogenic Imaging Spectrometer

For CRIS we are leveraging some of the most significant technology developments carried out for the International X-ray Observatory (IXO). We will use the same detector technology developed for the X-ray microcalorimeter spectrometer on IXO; for the optics we rely partially on the IXO silicon pore optics technology (SPO) and partially on classical Wolter I optics; and for the detector cooling we use a system which is functionally equivalent to the one proposed for IXO. The instrument has been optimized to give a high grasp (effective area x solid angle) by having a short focal length (2.5 m). Considering the maximum size of the cooled detector this corresponds to a 30″ field of view. With this configuration we maximize the effective area at lower energies (<1 keV) while retaining a reasonable effective area at 6 keV (>150 cm$^2$). The short focal length also has a low moment of inertia, necessary for fast re-pointing.

*Optics*: To meet the ORIGIN requirements of high effective area at 1 keV and a working range of 0.2–8 keV, we propose a telescope design based on a hybrid mirror technology: the outer part of the mirror will be built with Silicon Pore Optics (SPO) and the inner part with a Wolter type I telescope using standard Ni electroforming technology (figure 10). This choice reduces the mass (the SPO is very light). For the high-energy response the inner mirror is needed (as SPO optics has only been demonstrated at radii down to 0.3 m). SPO relies on using the flat surface of coated Si wafers as reflectors and stacking a set of these wafers in elements that, placed behind each other, can approximate the required geometry of a two-reflection focusing optical element. A fully automated stacking robot is operating to assemble stacks of plates according to the IXO optics design with 20 m focal length and a bending radius of 0.74 m. For our application we have to stack shorter wafers with an inner bending radius of 0.3 m. The smaller bending radius has been studied by finite element modeling and has been demonstrated by bending and bonding of two mirror plates with this radius. A Half Equivalent Width (HEW) angular resolution of <20" has been repeatedly demonstrated by X-ray pencil beam measurements performed at PTB in Berlin for a set up to 30 plates of an IXO mirror module, mounted in representative flight configuration. These results indicate that in the case of ORIGIN, the 30″ HEW requirement, being dominated by the conical approximation to the Wolter type I optics, is achievable. To increase the mirror reflectivity, the SPO will be coated with a three layer reflecting surface: C (25 Å) – Ni (25 Å) – Pt (300 Å). To estimate the effective area we assumed the proven thickness of the reflectors (170 μm) but FEM calculations suggest that we can reduce this to 120 μm resulting in a gain of 10% of the area. For the Wolter type I telescope the baseline is Au coating, but with using a Pt/C multi-layer coating (15 to 50 layers ranging between 40–100 Å using an extra Pt layer on the outside), we can boost the area at 6 keV by about 70%. The HEW increases from close to 20" in the center to about 40" near the edge of the field of view. For the higher energies the angular resolution can be as good as 15–20" in the center. The vignetting is modest at lower energies (less than 20% between the center and edge of the FoV) but increases for higher energies.

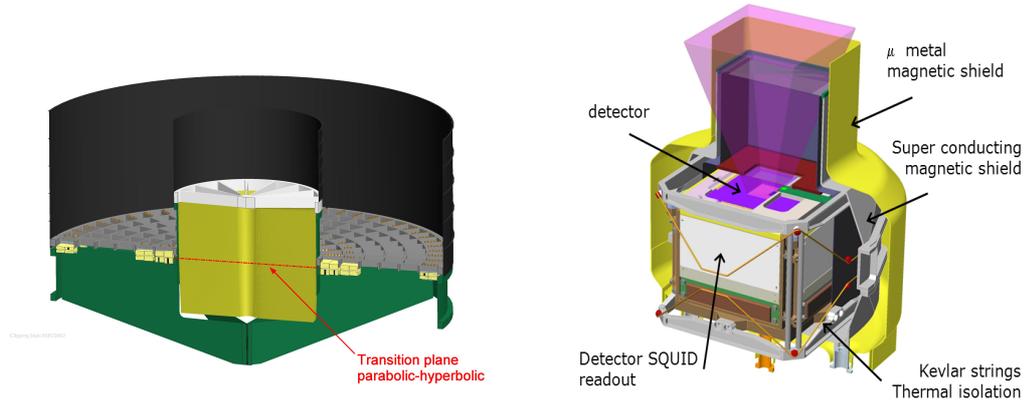

**Fig. 10** Left: Design of the CRIS hybrid mirror. Inner part: classical Ni formed shells mounted on a spider. Outer part: SPO modules (yellow) mounted in four petals (grey). The green part is the structural support. Right: engineering view of the CRIS focal plane assembly

*Detector:* For the detector we have selected an array of calorimeters for which the temperature rise after absorbing the photon, is measured by a very sensitive thermometer. Using a Transition Edge Sensor (TES) operated at ~100 mK a spectral resolution of <2.5 eV at 6 keV is feasible (Kilbourne et al. 2008, Gottardi et al. 2010). The detector assembly consists of the detector, its electronics and the cooling system. The detector has a number of different sections including an inner array of 26 x 26 pixels with energy resolution of 2.5 eV (optimized for $E_{max}$=10 keV), and an outer array of 72 x 72 pixels. In the outer array four pixels are read out by a single TES connected by 4 different strong thermal links. This type of detector allows identification of the X-ray absorbing pixel from the pulse shape before the TES and 4 absorbers come into thermal equilibrium (Smith et al. 2009). The resolution is <5 eV (optimized for $E_{max}$=5 keV). There is also a third array overlaying the outer array for detecting GRBs. This array consists of 20 x 26 pixels and is placed 8 mm out of focus. The intense X-ray beam of GRBs is spread over a sufficiently large number of pixels that the maximum count-rate capability of each pixel is not exceeded. The detector area of the outer array behind this GRB section, will of course, not be populated. The total number of channels to be read out is 2201. The focal plane assembly needs to be compact, and thermally and magnetically isolated from the environment (i.e. the temperature stability of the cold stage needs to better than 1 μK rms). The detector signals are multiplexed near the detector, which reduces the wiring between the cryogenic detector and room temperature. The SQUID read-out amplifiers are required to be in close proximity to the detectors, and these SQUIDs must also be magnetically well shielded. In figure 10 we show an engineering view of the focal plane assembly.

The cooling system for ORIGIN includes a combination of a last stage cooler (providing the 45 mK heat sink temperature for the detector), two Joule Thomson coolers (J-T) and four 2-stage Stirling Coolers. The JT coolers provide the 4K stage (each JT cooler is pre-cooled by 2-stage Stirling coolers) and the other 2-stage Stirling cooler cool the thermal shields. We have chosen a very conservative approach assuming that the dewar will be launched under vacuum (reducing the acoustic loads on the blocking filters). Tests on these filters are planned and, if successful, may eliminate the need for the vacuum enclosure, resulting in a mass saving (~50 kg). We have selected a combination of three adiabatic demagnetization refrigerators (ADR) to cool the detector assembly from 4 K down to 45 mK because of the high TRL level of this technology. Following magnetization of the salt pills, cooling is provided by the relaxation of the spins in the magnetized material. To reach the 45 mK level from 4 K the three ADRs (with 2 GLF stages and a single CPA salt pill for the last stage) need to be in series. The recycling time of the last stage cooler is less than 2 hours, with a hold time of 31 hours.

Table 1 Key characteristics of the Cryogenic Imaging Spectrometer (CRIS)

| Parameter | | Inner | Outer | GRB |
|---|---|---|---|---|
| Required (eV) | | 2.5 | 5.0 | 2.5 |
| Goal (eV) | | 1.5 | 3.0 | 1.5 |
| FoV (arcmin$^2$) | | 10 x 10 | 30 x 30 | 6 x 6 |
| | Full detector | | | |
| Energy range | 0.2–8 keV | | | |
| Angular resolution | 30" (HPD) | | | |
| Effective area | required (goal) | | | |
| @ 0.5 keV | 1000 (1500) cm$^2$ | | | |
| @ 1.5 keV | 700 (1000) cm$^2$ | | | |
| @ 6.0 keV | 100 (200) cm$^2$ | | | |
| Point source line detection sensitivity (5σ at 0.5 keV) | 2 10$^{-7}$ photons cm$^{-2}$ s$^{-1}$ | | | |
| Confusion limit for 0.5–2 keV | 10$^{-15}$ erg cm$^{-2}$s$^{-1}$ | | | |
| E-scale stability | 1 eV/hr | | | |
| Good grade events | >80% @ 50 counts/s/pix (ΔE nominal) | | | |
| Non X-ray background | 2 10$^{-2}$ counts/cm$^2$/s/keV | | | |

Fig. 11 shows the effective area of CRIS. The drop below 0.5 keV is mainly due to the optical filters in the cryostat. The edge around 2 keV is due to the mirror reflectivity and the drop at higher energies is a combination of the detector (absorber quantum efficiency for a given detector thickness) and a drop in the mirror effective area as a function of energy.

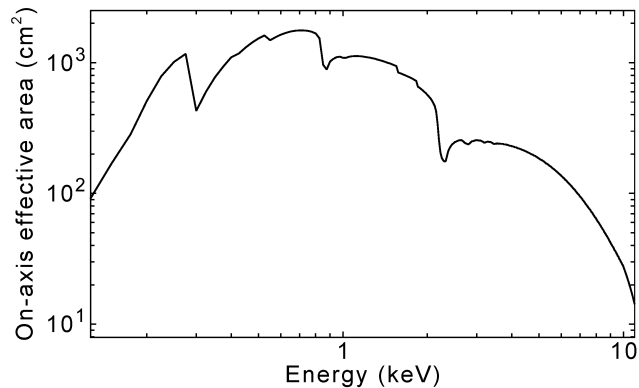

**Fig. 11** Estimated on-axis effective area for CRIS, as a function of energy.

## The Transient Event Detector

We will use a coded mask detector (Natalucci et al. 2007) to monitor a large fraction of the sky for transient events. The energy band between 5–200 keV is optimal to identify GRBs: the currently operational mission Swift works in this regime but with a higher threshold of 15 keV. For ORIGIN we have a modular approach with 4 units, tilted outward by ~30° on average. This yields a field of view as wide as 4 sr, with extension up to 4.6 sr for brighter bursts. The detector and mask size as well as the pixel size have been tuned to the instrument requirements (field of view, source location accuracy and sensitivity).

The design of TED is largely modular and based on the heritage of INTEGRAL and *Swift*. The 4 units are identical and the detection planes and masks have rectangular shapes. The detection plane of one instrument is an assembly of 12 CdZnTe (CZT) modules mounted on a spider structure. Each module in turn is formed by an array of 16x8 square crystals of size 1 cm$^2$ and thickness of 2 mm. An array of 4x4 pixels is then formed in each crystal by electrode segmentation. The detector pixel size is 2.5 mm and there are 2048 pixels/module. The active detection area is 1536 cm$^2$ for one unit. The coded mask has a pixel size of 2.8 mm, for a random pattern of holes and an open fraction of 40%. The mask is 0.8 mm thick and is fixed to a rectangular support grid on top of the shield. It will be built by etching or laser cutting single pieces of tungsten of 26.4x26.4 cm$^2$ each. On top of the mask we put a thin optical blocking filter to suppress the thermal load from partial exposure to sunlight. The angular resolution of this system is 23.8', which allows for a location accuracy of 3' (90% confidence) at the detection limit of 12σ.

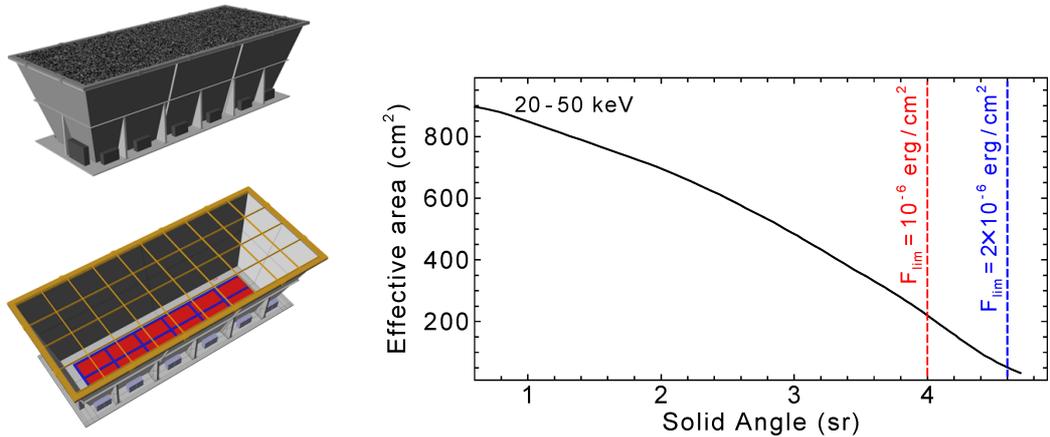

**Fig. 12** Left: a TED unit. The main parts are the detector modules, visible in red, the shield support and mask grid support. On the external side are the module bias boxes and the electronics. Right: the flux limit as function of solid angle covered by the TED. For reference we also provide the corresponding effective area (vertical axis). GRBs with a fluence of the prompt emission as low as 10$^{-6}$ erg/cm$^2$ occurring within 4 sr will be detected.

Each single module is equipped with a bias unit and a digital electronics board providing AD conversion of the signals. For each module specific bias voltages can be set. These bias boxes, providing power to the modules, are mounted on the unit close to each detector module and connected to it by flat cables as successfully implemented in the INTEGRAL/IBIS detector. For each unit, the 12 digital FEE boards are controlled by the Unit Electronics Box (UEB) mounted on the short side of the unit. This electronics provides configuration control (noisy pixels, low thresholds) and event processing. The TED Instrument Control Unit (ICU) receives data from the 4 UEBs and performs the TM/TC and S/C I/F functions, and the GRB trigger and positioning. An identical ICU is implemented as a cold redundant unit that remains switch-ed off in nominal conditions. The four TED units are tilted by 30° with respect to a plane that is inclined in turn by 15° from the satellite platform. The sensitivity of the instrument has been optimized and is consistent with the 4 sr coverage. The effective area is >200 cm$^2$ over a field of view of 4 sr, corresponding to a fluence limit prompt emission of 10$^{-6}$ erg/cm$^2$ (for a GRB of 20 s and a trigger integration time of 10 s). This is shown in figure 12. For bursts as bright as 2x10$^{-6}$ erg cm$^{-2}$ the field of view is even 4.6 sr.

Table 2 Key characteristics of TED

| Parameter | Value |
| --- | --- |
| Field of View | ~4 sr |
| Energy Range (keV) | 5–200 (3–200) |
| Angular Resolution | 23.8' |
| Source Location Accuracy (12σ) | <3' (goal: 2') |
| Energy res. (100 keV, FWHM) | ~3% |
| Count rate/unit (min/max) | 4–15 k |
| Sensitivity (ph cm$^{-2}$ s$^{-1}$ in 10 s, 5–150 keV) | 0.4 (12σ) |
| Software processing time (s) | 20 (10) |

During normal operations the TED units will monitor the sky waiting for a GRB or a transient to occur. In the meanwhile, TED will produce spectral-imaging data (*normal mode)* in which detector images are provided in a pre-defined set of energy bands. The on-board trigger will be based on sampling count rates at unit and module levels with different integration times and energy ranges, and includes an imaging trigger based on an on-board catalog of known sources, to discriminate non-GRB events. This procedure is standard in satellites like Swift and AGILE. When the trigger signal is produced, the ICU performs the imaging part of the analysis (*trigger mode*) and the GRB position is generated and transmitted to the S/C within ~20 s. It is not required to collect data during slews. For safety TED units will be put autonomously in a safe mode when the Sun is in their field of view. While this happens for one (or two) units the other TEDs continue to operate nominally.

### The Burst InfraRed Telescope

In order to ensure the discovery of the optical counterpart of a GRB, pin-pointing its host galaxy and determining the redshift, an optical/near-IR telescope is planned. To date, all GRB redshifts have been measured with optical or near-IR instruments. Determination of GRB redshifts from zero to twelve is achieved with a wavelength range of 0.5 to 1.7 μm. BIRT will detect the GRB counterpart and measure the redshift by using a combination of multi-band imaging (IMAGE), R=1000 integral-field dispersed spectroscopy (HIGH-RES) and R=20 low dispersion slitless spectroscopy (LOW-RES).

Initially, in parallel with CRIS observations, the counterpart will be observed in the LOW-RES mode, which will identify high-redshift GRBs and provide an approximate redshift from the Lyman-alpha break, while they are in their brightest phase. Once CRIS has identified the X-ray counterpart, the precise position will be determined in the IMAGE mode. For brighter counterparts, HIGH-RES spectroscopy will measure a precise redshift from absorption lines.

BIRT is an optical to IR Cassegrain telescope with a 0.7 m primary mirror of Silicon Carbide with a Hawaii-2RG 2K x 2K HgCdTe detector. Below the primary mirror the image plane is split into 3 independent optical paths, splitting the beam over the 3 BIRT operating modes. In the LOW-RES mode the dispersing element is a prism. A 4 x 4 spatial-pixel image slicer directs the light onto a grating for the HIGH-RES mode. The IMAGING mode uses dichroics to separate the 4 different bands, which are imaged simultaneously. Focus is maintained by heaters on metering rods between the primary and secondary mirrors. The telescope utilizes a baffle tube which extends 3.8 m forward of the primary mirror, together with a secondary mirror baffle, an inner-primary mirror baffle; baffling of the field and pupil stops within the instrument box for straylight suppression. With this design the observing constraints of GRBs will be dictated by the CRIS instrument. The key characteristics are given in table 3. The telescope will be passively cooled to ~270K. The detector and the camera optical baffle must be cooled actively to minimize the background signal. This requires a miniature pulse-tube cooler (MPTC) connected to a radiator.

Table 3 Prime characteristics of BIRT

| | IMAGING | LOW-RES | HIGH-RES |
|---|---|---|---|
| Wavelength range (μm) | 0.5–0.7, 0.7–1.0, 1.0–1.3, 1.3–1.7 | 0.7–1.7 | 0.5–1.7 |
| Field of View | 1'x1' | 6'x6' | 2.1"x2.1" |
| Limiting mag. (AB) | H=22.2 | H=20.8 | H=19.3 |
| Spectral resolution R | 3–4 | 20 | 1000 |
| Spatial resolution | 0.2", 0.3", 0.4", 0.5" | 0.3–0.5" | 0.2–0.5" |

The instrument will gather data from 3 distinct sky regions on different parts of the detector. A 6'x6' field of view is used for low-resolution slitless spectroscopy with R~20. A section of 1'x1', slightly offset with respect to the LOW-RES field of view, is used for imaging in 4 bands. A small region of 2"x2" is used for high-resolution integral field spectroscopy with a resolution of R=1000. These distinct regions are mapped onto a single detector. Targets are switched between LOW-RES and IMAGE by re-pointing the satellite using the GRB position as provided by CRIS. For the HIGH-RES mode the GRB position inside the BIRT field of view must be known with an accuracy of 0.1". This is obtained from the data in the IMAGE mode. Placement of the GRB on the HIGH-RES section and corrections for drift and high frequency perturbations (e.g. due to the compressors of the cryo-cooler) is achieved by a tip/tilt fold mirror internally to the BIRT. Slow drifts of the spacecraft will be monitored using stars in the IMAGE section and high frequency perturbations will be monitored using a gyro directly attached to BIRT.

## Spacecraft

For the spacecraft design we take a very conservative approach. In practice this implies that we seek to decouple spacecraft systems as much as possible, neglecting any potential benefits from a closer integration. This approach minimizes system complexity, taking advantage of the large mass capability for a satellite in LEO with a Soyuz launcher. This increases the weight, but the simpler design, based on off-the-shelf units, will reduce the cost. In figure 13 we show an engineering view of the satellite. For most subsystems (power, thermal subsystem, data handling, propulsion and deorbiting, mission operations) we selected off- the-shelve systems. For the communications we require the X-band. In addition, satellite-to-relay satellite communication is included for fast transfer of GRB positions to the ground to enable rapid follow up. The AOCS system (fast repointing) and the long baffles to reduce straylight have been studied in detail.

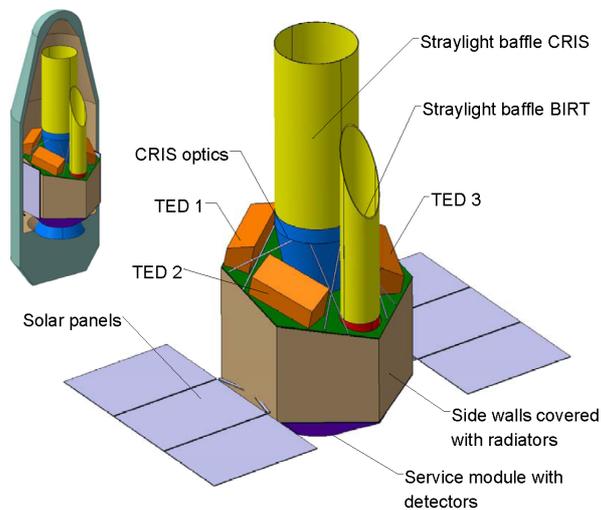

**Fig. 13** Accommodation of the payload and satellite systems in ORIGIN and accommodation in the fairing (top left). The launch mass, including adaptor and margins is 2904 kg and the total power including margin is 2790 W.

*Attitude and orbit control*: The attitude and orbit control has to meet the requirements for a fast responding satellite: (a) 3 'slow' re-pointings per orbit (to allow for a 70% observing efficiency in a LEO); (b) one fast, autonomous, repointing per orbit; (c) each orientation fixed with respect to a celestial object and (d) illumination due to Sun and Earth to be avoided along the telescope bore sights The mastering of fast repointing AOCS solutions within Europe in the last few years (e.g. Pleiades) ensures that these challenges can be met. To reduce the required capability of the actuators the Moment of Inertia (MoI) can be minimized by symmetric mounting of equipment near the Center of Mass (CoM). A major driver is the focal length of CRIS (i.e. the distance to the heavy mirror). This is a trade-off between the actuator sizing and the higher energy response of the telescope. For this we rely on existing and proven technology although mass savings could be achieved if the bending radius of the SPO can be reduced from 0.3 m to 0.2 m. The CoM and MoI for the satellite are (0, 0, 1.2) m with respect to the launch adaptor and (5500, 5300, 2400) kg.m$^2$, respectively. To achieve fast repointing we have baselined Control Momentum Gyros (CMGs) which give an overall lower system mass and power than a Reaction Wheel solution and allow the 1°/s agility to be met using the of-the-shelf Honeywell M-50 in a 4 unit pyramid configuration. For the unloading of the CMG momentum we will use Magnetic Torquers during every orbit (noon time). The AOCS system is complemented by 2 Sun sensors, 2 GPS systems, 1 IMU (Inertial Measurement Unit) and 2 magnetometers. As star tracker we have selected the SODERN Hydra system. All units chosen are available off-the-shelf.

*Straylight*: In a Low Earth Orbit not only the repointing is important but also the sky visibility, as the instruments (CRIS and BIRT) are sensitive to straylight (<2.5 10$^9$ photon/cm$^2$/s at the dewar entrance for CRIS and <10$^5$ photons/cm$^2$/s on the BIRT detector). Based on a preliminary analysis we need a reduction of the reflected Sun/Earth light of 10$^{-6}$ for CRIS, corresponding to at least 2 reflections inside the baffle. An estimate of the self-baffling properties of the X-ray optics including integrated baffles indicates that a reduction of 10$^{-3}$ from the optical straylight baffle is sufficient. With a 3.5 m long baffle for CRIS (yellow tube figure 13) this is feasible with an exclusion cone of half angle 24°. The baffle length can be traded against the fraction of sky available but the baseline length fits comfortably within the fairing whilst giving access to >53% of the sky during the whole orbit. For BIRT attenuation by 10$^{10}$ is required. This is achieved by a long baffle tube together with baffles on the secondary, around the M1 Cassegrain aperture, and within the instrument enclosure. The primary baffle tube is lined with conical vanes such that the majority of the incident light not absorbed is directed back into space. The baffle tube has a sloping entrance so that it is longer on the CRIS side to prevent light scattering from the outside of the CRIS baffle into the BIRT optics. The BIRT baffle length (on the shortest edge) to aperture ratio is larger than that used on Swift UVOT (5.4 for BIRT, compared with 5.0 for UVOT), and BIRT incorporates field and pupil stops within its optical trains so that the BIRT straylight rejection will be superior to that of UVOT, which has demonstrated near-zodiacal-limited performance down to an Earth limb angle of 25° (Breeveld et al. 2010).

## Science operations

ORIGIN will follow the typical ESA share of responsibilities: ESA is responsible for the Mission Operation Control (MOC) and the Science Operations Center (SOC). The SOC sets the observing schedule and take advantage of the capability to autonomously repoint the satellite. A list of pre-planned targets will be uploaded, allowing the independent execution of the schedule for several days. For the guest observer program the target selection will be carried out through an open call to the community and subsequent peer review. Key programs, requiring long observations, will be selected in consultation with the community through the Science Working Team (representatives of instrument teams and the science community).

A Science Data Center (SDC) will be established to routinely process flight data and to provide the analysis software to the community. Responsibilities of the SDC include data processing, integration of instrument specific software in the analysis package (including testing), delivery of generic tools for the analysis of the science data and the distribution of this software to the community including user support for the use of these tools. The instrument teams will be responsible for the health and calibration of their instruments, for defining the trigger criteria and providing the instrument specific software. Data from valid GRB triggers and consequent follow up measurements by the other ORIGIN instruments (positions, spectra, light curves) will be transmitted to the ground in real time and distributed via the internet to the worldwide community. The data will be inspected on a daily basis by instrument teams and the SDC to ensure quality. This reduces the load on the SOC. This working scheme operates successfully for *Swift*.

# Conclusion

ORIGIN is an exciting mission to study the metal enrichment from redshifts > 7 up to the present. We selected some of the prime science which demonstrates the power of high spectral resolution observations in the soft X-ray band to study cosmic chemical evolution. With a five year mission duration we expect around 65 GRBs at redshifts > 7. This requires a satellite which identifies and localizes bursts and can autonomously slew to the position of the GRB. For bursts with low metallicity, the redshift cannot be determined from the absorption lines in the X-ray spectra and therefore the payload includes the capability to determine the redshift in the IR. This payload is well feasible within the envelope of an M3 mission. Detailed studies have demonstrated that fast re-pointing with a large sky visibility (> 53%) is feasible in a Low Earth Orbit using available technology. Also the other satellite systems are not over demanding. For the instruments we use available technology (the coded mask instrument to detect GRBs and the IR telescope to determine independently the redshift) or exploit technology which has been developed for IXO (the cryogenic instrument and the optics).

# Acknowledgements

The team likes to express its appreciation for the support of Astrium UK for the present study. Earlier studies, which also confirmed the feasibility of this concept were carried out by Thales/Alenia and NASA/MSFC.

# References

Alvarez, M. A., Busha, M., Abel, T., Wechsler, R. H.: Connecting Reionization to the Local Universe. ApJ 703, L167 (2009)

Balestra, I., Tozzi, P., Ettori, S., Rosati, P., Borgani, S., Mainieri, V., Norman, C.: Evolution in the Iron Abundance of the ICM. P Th PS 169, 49 (2007)

Bastian, N., Covey, K. R., Meyer, M. R.: A Universal Stellar Initial Mass Function? A Critical Look at Variations. ARA&A 48, 339 (2010)

Breeveld, A. A., Curran, P. A., Hoversten, E. A., et al.: Further calibration of the Swift ultraviolet/optical telescope. MNRAS 406, 1687 (2010)

Branchini, E., Ursino, E., Corsi, A., et al.: Studying the Warm Hot Intergalactic Medium with Gamma-Ray Bursts. ApJ 697, 328 (2009)

Borgani, S., Murante, G., Springel, V., et al.: X-ray properties of galaxy clusters and groups from a cosmological hydrodynamical simulation. MNRAS 348, 1078 (2004)

Boyarsky, A., Ruchayskiy, O., Shaposhnikov, M.: The Role of Sterile Neutrinos in Cosmology and Astrophysics. ARNPS 59, 191 (2009a)

Boyarsky, A., Lesgourgues, J., Ruchayskiy, O., Viel, M.: Realistic Sterile Neutrino Dark Matter with KeV Mass does not Contradict Cosmological Bounds, PhRvL 102, 201304 (2009b)

Bromm, V., Loeb, A.: GRB Cosmology: Probing the Early Universe. AIPC 937, 532 (2007)

Butler, N. R., Bloom, J. S., Poznanski, D.: The Cosmic Rate, Luminosity Function, and Intrinsic Correlations of Long Gamma-Ray Bursts. ApJ 711, 495 (2010)

Bykov, A. M., Paerels, F. B. S., Petrosian, V.: Equilibration Processes in the Warm-Hot Intergalactic Medium. SSRv 134, 141 (2008)

Campana, S., Romano, P., Covino, S., et al.: Evidence for intrinsic absorption in the Swift X-ray afterglows. A&A 449, 61 (2006)

Cen, R., Ostriker, J. P.: Where Are the Baryons? II. Feedback Effects. ApJ 650, 560 (2006)

Chieffi, A., Limongi, M.: Explosive Yields of Massive Stars from $Z = 0$ to $Z = Z_{solar}$. ApJ 608, 405 (2004)

de Plaa, J., Werner, N., Bleeker, J. A. M., Vink, J., Kaastra, J. S., Méndez, M.: Constraining supernova models using the hot gas in clusters of galaxies. A&A 465, 345 (2007)

Diemand, J., Kuhlen, M., Madau, P.: Clumps and streams in the local dark matter distribution. Nature 454, 735 (2008)

Fan, X., Carilli, C. L., Keating, B.: Observational Constraints on Cosmic Reionization. ARA&A 44, 415 (2006)

Gabriel, A. H., Phillips, K. J. H.: Dielectronic satellite spectra for highly charged helium-like ions. IV - Iron satellite lines as a measure of non-thermal electron energy distributions. MNRAS 189, 319 (1979)

Gallerani, S., Salvaterra, R., Ferrara, A., Choudhury, T. R.: Testing reionization with gamma-ray burst absorption spectra. MNRAS 388, L84 (2008)

Gonzalez, A. H., Zaritsky, D., Zabludoff, A. I.: A Census of Baryons in Galaxy Clusters and Groups. ApJ 666, 147 (2007)

Gottardi, L., et al. 2010, Proc. ASC, in press

Greiner, J., Krühler, T., Klose, S., et al.: The nature of ``dark'' gamma-ray bursts. A&A 526, A30 (2011)

Heger, A., Fryer, C. L., Woosley, S. E., Langer, N., Hartmann, D. H.: How Massive Single Stars End Their Life. ApJ 591, 288 (2003)

Heger, A., Woosley, S. E.: Nucleosynthesis and Evolution of Massive Metal-free Stars. ApJ 724, 341 (2010)

Jakobsson, P., Levan, A., Fynbo, J. P. U.: A mean redshift of 2.8 for Swift gamma-ray bursts. A&A 447, 897 (2006)

Kawai, N., Yamada, T., Kosugi, G., Hattori, T., Aoki, K.: GRB 050904: Subaru optical spectroscopy. GCN 3937, 1 (2005)

Kann, D. A., Klose, S., Zhang, B., et al.: The Afterglows of Swift-era Gamma-ray Bursts. I. Comparing pre-Swift and Swift-era Long/Soft (Type II) GRB Optical Afterglows. ApJ 720, 1513 (2010)

Kilbourne, C. A., Doriese, W. B., Bandler, S. R.: Multiplexed readout of uniform arrays of TES X-ray microcalorimeters suitable for Constellation-X. Proc. SPIE 7011, 701104 (2008)

King, A., O'Brien, P. T., Goad, M. R., Osborne, J., Olsson, E., Page, K.: Gamma-Ray Bursts: Restarting the Engine. ApJ 630, L113 (2005)

Kravtsov, A. V., Vikhlinin, A., Nagai, D.: A New Robust Low-Scatter X-Ray Mass Indicator for Clusters of Galaxies. ApJ 650, 128 (2006)

Linder, E.V.: A Cosmic Vision Beyond Einstein. Proc. of "Identification of dark matter 2008". August 18-22, 2008, Stockholm, Sweden, 42 (2008)

Loeb, A., Ferrara, A., Ellis, R. S.: First Light in the Universe. Springer (2008)

Maiolino, R., Nagao, T., Grazian, A., et al.: AMAZE. I. The evolution of the mass-metallicity relation at $z > 3$. A&A 488, 463 (2008)

McCammon, D., Almy, R., Apodaca, E., et al.: A High Spectral Resolution Observation of the Soft X-Ray Diffuse Background with Thermal Detectors. ApJ 576, 188 (2002)

Natalucci, L.; Feroci, M.; Quadrini, E., et al.: Design of a CZT gamma-camera for GRB and fast transient follow-up: a wide-field-monitor for the EDGE mission. Proc. SPIE 6686, 66860T (2007)

Nicastro, F., Zezas, A., Drake, J., et al.: Chandra Discovery of a Tree in the X-Ray Forest toward PKS 2155-304: The Local Filament? ApJ 573, 157 (2002)

Paerels, F.B.S., Kaastra, J., Ohashi, T., Richter, P., Bykov, A., Nevalainen, J.: Future Instrumentation for the Study of the Warm-Hot Intergalactic Medium. SSRv 134, 405 (2008)

Primack, J. R., Gilmore, R. C., Somerville, R. S.: Diffuse Extragalactic Background Radiation. AIPC 1085, 71 (2008)

Schindler, S., Diaferio, A.: Metal Enrichment Processes. SSRv 134, 363 (2008)

Salvaterra, R., Campana, S., Chincarini, G., Covino, S., Tagliaferri, G.: Gamma-ray bursts from the early Universe: predictions for present-day and future instruments. MNRAS 385, 189 (2008)

Salvaterra, R., Della Valle, M., Campana, S., et al.: GRB090423 at a redshift of $z\sim8.1$. Nature 461, 1258 (2009)

Savaglio, S., Fall, S. M.: Dust Depletion and Extinction in a Gamma-Ray Burst Afterglow. ApJ 614, 293 (2004)

Schady, P., Savaglio, S., Krühler, T., Greiner, J., Rau, A.: The missing gas problem in GRB host galaxies: evidence for a highly ionised component. A&A 525, A113 (2011)

Schneider, R., Ferrara, A., Salvaterra, R., Omukai, K., Bromm, V.: Low-mass relics of early star formation. Nature 422, 869 (2003)

Smith, S.J., Bandler, S.R., Brekosky, R.P., et al.: Development of Position-Sensitive Transition-Edge Sensor X-Ray Detectors. IEEE Trans. Appl. Superc. (2009). doi: 10.1109/TASC.2009.2019557

Snowden, S. L., Egger, R., Freyberg, M. J.: ROSAT Survey Diffuse X-Ray Background Maps. II. ApJ 485, 125 (1997)

Soderberg, A., Grindlay, J. E., Bloom, J. S.: The Dynamic X-ray Sky of the Local Universe. Astro2010, 278 (2009)

Springel, V., Frenk, C. S., White, S. D. M.: The large-scale structure of the Universe. Nature 440, 1137 (2006)

Takei, Y., Ursino, E., Branchini, E., et al.: Studying the Warm-Hot Intergalactic Medium in Emission. ApJ submitted (2010)

Ursino, E., Branchini, E., Galeazzi, M., et al.: Expected properties of the Two-Point Autocorrelation Function of the IGM. MNRAS accepted (2010)

Viel, M., Haehnelt, M. G., Springel, V.: Inferring the dark matter power spectrum from the Lyman α forest in high-resolution QSO absorption spectra. MNRAS 354, 684 (2004)

Werner, N., de Plaa, J., Kaastra, J. S.: XMM-Newton spectroscopy of the cluster of galaxies 2A 0335+096. A&A 449, 475 (2006)